\documentstyle[amsfonts,epsfig,11pt]{article}

\def\mypagenumber{1}

\def\myend{\end{document}}

\normalsize

\newcounter{sxn}

\newcounter{axn}

\date{}

\newdimen\mybaselineskip
\newcommand{\beeq}{\begin{equation}}
\newcommand{\eneq}{\end{equation}}
\newcommand{\be}{\begin{eqnarray}}
\newcommand{\ee}{\end{eqnarray}}
\newcommand{\bpic}{\begin{picture}}
\newcommand{\epic}{\end{picture}}

\def\la{\raise.16ex\hbox{$\langle$} \, }
\def\ra{\, \raise.16ex\hbox{$\rangle$} }

\def\psibar{ \psi \kern-.65em\raise.6em\hbox{$-$} }
\def\mbar{ m \kern-.78em\raise.4em\hbox{$-$}\lower.4em\hbox{} }

\def\n@space{\nulldelimiterspace=0pt \mathsurround=0pt }
\def\huge#1{{\hbox{$\left#1\vbox to 20.5pt{}\right.\n@space$}}}

\def\myskip{\noalign{\kern 8pt}}
\def\myeqspace{\noalign{\kern 10pt}}

\def\boxit#1{$\vcenter{\hrule\hbox{\vrule\kern3pt
    \vbox{\kern3pt\hbox{#1}\kern3pt}\kern3pt\vrule}\hrule}$}
\def\bigbox#1{$\vcenter{\hrule\hbox{\vrule\kern5pt
     \vbox{\kern5pt\hbox{#1}\kern5pt}\kern5pt\vrule}\hrule}$}

\def\ignore#1{{}}

\begin{document}

\bibliographystyle{unsrt}
\footskip 1.0cm

\thispagestyle{empty}
\setcounter{page}{\mypagenumber}


\begin{flushright}{BRX-TH-520\\}
\end{flushright}

\vspace{2.5cm}
\begin{center}
{\LARGE \bf {Shortcuts to 
high symmetry solutions in gravitational theories }}\\
\vspace{2cm}
{\large S. Deser \hskip 0.3 cm and \hskip 0.3 cm
Bayram Tekin\footnote{e-mail:~
deser,tekin@brandeis.edu}~\footnote{Address after July 15, 
{\it{Physics Department, Middle 
East Technical University, 06531, Ankara, Turkey}.} }}\\
\vspace{.5cm}
{\it {Department of Physics, Brandeis University, Waltham, MA 02454,
USA}}\\

\end{center}

\vspace*{1.5cm}


\begin{abstract}
\baselineskip=18pt
We apply the Weyl method, as sanctioned by Palais' symmetric criticality 
theorems, to obtain those -highly symmetric -geometries  amenable to 
explicit solution, in generic gravitational models and dimension. The 
technique consists of judiciously violating the rules of variational 
principles by inserting highly symmetric, and seemingly gauge fixed, 
metrics into the action, then varying it directly to arrive at a small 
number of transparent, indexless, field equations. Illustrations 
include  spherically and axially symmetric solutions in a wide range of 
models beyond $D=4$ Einstein theory; already at $D=4$, novel results 
emerge such as exclusion of Schwarzschild solutions in cubic 
curvature models and restrictions on ``independent'' integration parameters 
in quadratic ones. Another application of 
Weyl's method is an easy derivation of Birkhoff's theorem in systems with 
only tensor modes. Other uses are also suggested.

\end{abstract}
\vfill

 
\newpage



\normalsize
\baselineskip=24pt plus 1pt minus 1pt
\parindent=25pt

\section{Introduction}

In his classic monograph on general relativity \cite{weyl},
Weyl derived the Schwarzschild metric by a very elegant method 
leading to just two scalar field equations instead of the usual 
tensorial set. The simplification was achieved by choosing 
the input metric not only to exhibit spherical symmetry and 
{\it{a priori}} time-independence but also to select a particular gauge, 
Schwarzschild coordinates; in our (slightly different) parametrization,
\be
ds^2 = -a(r)b(r)^2\,dt^2 + {dr^2 \over a(r) } + r^2 d\Omega_2
\label{metric1},
\ee
where $ d\Omega_2$ is the 2-sphere element, the Einstein action simply 
takes the form
\be 
I = \int d^4x \sqrt{-g} R \rightarrow \int_0^\infty\,dr \, r(a-1)b'   
\label{action1}.
\ee
The two Euler-Lagrange equations obtained by varying the 
dependent variables $(a, b)$ in (\ref{action1}) 
are not only of first derivative order, 
but trivially integrable to the correct metric,
$a= 1-c_1/r$, $b= c_2$. It did not escape subsequent 
readers that this procedure 
seems to violate all the tenets of variational principles: 
It inter-compares solely static~\footnote{ This 
assumption is also tacit and 
unjustified in many textbooks.}, spherically 
symmetric geometries and even specifies the gauge! The correctness of 
the results, however, suggests that 
that there must be something to this procedure. Indeed, 
a mere fifty years later, a mathematical justification (
and equally important, its limitations) was given  by the 
symmetric criticality theorems of Palais \cite{palais}. 
We will not expound them here.~\footnote{For a recent discussion in the 
framework of GR, see \cite{torre}. }

Surprisingly, there has been very little application of 
this streamlined approach to GR, let alone to more complicated 
gravitational models involving higher curvature powers and generic 
dimension. Our aim here is the modest one of advertising the virtues of 
this technique through concrete generation of spherically and axially 
symmetric solutions in a variety of models, including matter and 
various string-generated corrections to Einstein theory such as 
the conformal Weyl action in $D=4$, higher powers of curvature ,
or Gauss-Bonnet terms in $D > 4$. We mention several novel results: 
pure $R^2$ gravity in $D=4$ allows either Schwarzschild-deSitter (SdS) 
or Reissner-Nordstrom (RN) solutions but not both; the actions 
cubic in the Weyl tensor plus any lower power of curvature 
at $D=4$ do not permit the Schwarzschild metric; in the four 
dimensional $R_{\mu \nu\rho \sigma}^2$ model, the only solution of the 
form $-g_{00} = g^{rr}$ is SdS. 

We hope that 
further novel applications will be found. One domain is obviously 
that of `brane' metrics. Another would exploit the fact that, under 
precisely defined circumstances, it is possible to generate correct 
solutions-along with metrics that are {\it{not}} correct-despite even 
more egregious violations of the ground rules; since it is easy 
to screen out non-solutions, one may hope to thereby obtain as 
yet unknown geometries in more complicated models. 
Separately, we will obtain the Birkhoff theorem, for 
models without scalar or vector modes, as another type of application .
Also, depending how one chooses to parametrize the metric,
one might generate solutions in other coordinates, although care 
is required here. More generally, we caution that the fine print 
must always be read with care.

\section{Spherical Symmetry}
 
The most general spherically symmetric metric in dimension 
$D$ and Schwarzschild gauge is
\be
ds^2 = -a(r,t)b(r,t)^2 dt^2 + {dr^2\over a(r,t)} + r^2d\Omega_{D-2}
\label{metric2},
\ee
where $d\Omega_{D-2}$ is the $D-2$ sphere element and the 
two functions $(a,b)$ depend on $(r,t)$ {\it{a priori}}. 
This ``gauge'' is well-known to be very special in that it is not so 
much a gauge choice but rather a natural expression of the symmetry 
orbits \cite{palais}. 
We have already noted its advantage in GR of reducing 
the derivative order of the action ({\ref{action1}).
~\footnote{ An example of illegal gauge choice is to further reduce 
the $(r,t)$ 2-space
to a single function, say to conformally flat form with 
$t$-independence. We know that the Schwarzschild metric 
cannot be so expressed. } 

Let us start with the most elementary example of 
cosmological GR, 
\be
I = \int d^D x \sqrt{-g}( R + \Lambda ).
\label{actioncosmo}
\ee
The scalar curvature and volume form are respectively 
\be
R&&= a'' +{2ab''\over b} + {3a'b'\over b} + 2(D-2){1\over r}(a' +{ ab'
\over b}) + (D-2)(D-3){1\over r^2}(a-1) \\ \nonumber
&&+{1\over a^2 b^2}(\ddot{a} - 2{\dot{a}^2\over a} - {\dot{a}\dot{b}\over b}), 
\hskip 1 cm \sqrt{-g} = b r^{D-2}.
\label{scalar}
\ee
The time derivatives all drop out in the action (\ref{actioncosmo} )
\be 
I = (D-2)\int_0^\infty\,dr\,r^{D-3}(a-1-{\Lambda r^2 
\over D-2 })b';    
\label{cosmological}
\ee
absence of any time derivatives in the integrated 
form is Birkhoff's theorem. 
It is now also obvious that the solution of 
(\ref{cosmological}) is SdS,
\be
a(r) =  1 + {c_1 \over r^{D-3}} + {\Lambda \over D-2}r^2, \hskip 2 cm 
b(r) = c_2
\ee
for $D >3$, and the conical SdS
\be
a = c_1 +\Lambda r^2, \hskip  2cm b = c_2,
\ee
for $D=3$. This agrees of course with the known results \cite{jackiw,btz}.

We can also add a Maxwell term, $ -{1 \over 4} \int \sqrt{-g}F_{\mu \nu}^2$, 
to the Einstein action to obtain the RN solution. 
A spherically symmetric and static (Birkhoff's theorem is also immediate 
here) gauge field can be written as 
$A = A_0(r)dt$, which then leads to the reduced cosmological 
Einstein-Maxwell system: 
\be
I = \int_0^\infty\,dr\,\left \{ r^{D-3}(a-1-{\Lambda r^2 
\over {D-2}})b' - {1\over 2 b}r^{D-2}{A_0'}^2  \right \}.  
\ee
Varying the action with respect to $A_0$, one obtains
\be
A_0 =  {b(r) c_3 \over r^{D-3}}.
\ee
The remaining equations follow easily and lead to RNdS: 
\be
b(r) = c_2, \hskip 2 cm 
a(r) =  1 + {c_1 \over r^{D-3}} + {\Lambda \over D-2}r^2 + 
{q^2 \over r^{2D-6}}.
\ee

Next, we run through some other 
gravitational models of current interest.

{\bf{ I  Birkhoff Theorem in Einstein-Gauss-Bonnet (EGB) theory }}

The EGB model frequently arises in the gravity and string theory literature.
Its spherically symmetric solutions were given in \cite{boulware} using 
in fact the Weyl trick, but assuming $t$-independence from the outset.
Here, we derive the latter. The action 
\be
I =  \int d^D x \sqrt{-g} \left \{ R + \alpha ( R^2 - 4R_{\mu \nu}^2 + 
R_{\mu \nu \sigma \rho}^2  )\right \}    
\ee
reduces to the following simple form for the metric (\ref{metric2})
\be
I = \int_0^\infty dr r^{D-3}(a-1)b'\left \{ 1+ \tilde{\alpha}r^{-2}(a-1)
\right \},
\ee
where $\tilde{\alpha}$ vanishes for $D=4$ and otherwise is proportional 
to $\alpha$. As in the case of pure Einstein theory, the action 
indeed has no time derivatives. The explicit solution of \cite{boulware}
\be
a(r) = 1 +{ r^2\over 2\tilde{\alpha}} \pm 
{r^2\over 2\tilde{\alpha}}\sqrt{ 1 + {c_1\tilde{\alpha} \over r^{D-1}}},
\ee
then follows directly. To complete our discussion of EGB theory, 
let us add the Maxwell term,
\be
I = \int_0^\infty dr \left \{ r^{D-3}(a-1)b' 
[ 1+ \tilde{\alpha}r^{-2}(a-1) ] -{1\over 2 b} r^{D-2}{A_0'}^2 \right \}.
\ee
Then the ``RNEGB'' solution reads : 
\be
b(r) = c, \hskip 2 cm 
a(r) = 1 +{ r^2\over 2\tilde{\alpha}} \pm 
{r^2\over 2\tilde{\alpha}}\sqrt{ 1 + {c_1\tilde{\alpha} \over r^{D-1}} + 
{q^2\tilde{\alpha} \over r^{2D-8} }}.
\ee

{\bf {II $C^2$ and $C^3$ models }} 

Explicit spherically symmetric solutions of conformal, $C^2$, gravity  
were given in \cite{riegert, mannheim}. Birkhoff's theorem was also
demonstrated in \cite{riegert}. Here, we show how the Weyl trick provides 
us with a shortcut to the explicit solution. Dropping time dependence 
for simplicity, the static metric given in (\ref{metric1}) reduces the action
\be
I=\int d^4 x \sqrt{-g} C_{\mu \nu \rho \sigma} C^{\mu \nu \rho \sigma }
\ee
to the 1-dimensional form
\be
I = \int_0^\infty dr{1\over  b r^2} 
\left \{ r^2 b a'' + 3r^2 a'b' + 2r^2 a b'' - 2rba' 
-2arb' - 2b + 2ba \right \}^2.
\label{weylaction}
\ee
After varying (\ref{weylaction}) with respect to $(a, b)$, 
one can set $b=c$ and integrate 
one of the two equations to obtain
\be
a(r) =  c_1 + {c_2 \over r} + c_3 r +  c_4 r^2, 
\ee  
The remaining equation then imposes the following condition on the 
integration constants,
\be
1 - c_1^2 + 3c_2 c_3 = 0. 
\label{condition}
\ee
The above solution seemingly accommodates both  
the Schwarzschild and the linear term $c_3 r$: However, 
if both were present  
simultaneously, the condition (\ref{condition}) would imply that 
$c_1 \ne 1$, namely a conical singularity (and not merely a horizon).
Thus, one must set $c_1 = 1$ to avoid the singularity and 
impose the stronger version of the condition (\ref{condition}), thereby  
excluding either the linear potential term or $1/r$.

Let us next sketch the spherically symmetric solutions of the 
cubic conformal curvature model, 
\be
I =\int d^4 x \sqrt{-g} C_{\mu \nu \rho \sigma}C^{\mu \nu}\,_{\alpha \beta} 
C^{\alpha \beta \rho \sigma },    
\ee
which, for the metric (\ref{metric1}), reduces to
\be
I = \int_0^\infty dr{1\over  b^2 r^4} 
\left \{ r^2 b a'' + 3r^2 a'b' + 2r^2 a b'' - 2rba' 
-2arb' - 2b + 2ba \right \}^3.
\ee
Once again, after varying the action with respect to $(a, b)$ and 
setting~\footnote{ In Einstein theory, $b=c$ is a consequence of the 
field equations. Here and below, we select this class without checking 
whether it is still forced. In either case, the metrics so obtained 
are still solutions of course. } $b = c$, one can show that there can 
be no $M/r$ term in the 
metric: The Schwarzschild metric is {\it{not}} a solution. Instead, 
one can see that $a(r) = 1 + c_1 r + c_2 r^2$, solves the 
equations: It yields, as expected,  
the most general conformally flat metric (in the spherical coordinates).  
To our knowledge, this is the simplest gravitational model in $D=4$ 
that actually excludes Schwarzschild horizons, an exclusion that also 
extends to the generic $R+ R^2 + C^3$ cases as well, since the other 
terms vanish for Schwarzschild metrics.   

{\bf{ III Solutions of pure $R^2$ theory }}

Let us consider the action 
\be
&&I = \int d^D x \sqrt{-g} R^2 \\ \nonumber
&&\rightarrow \int dr r^{D-2}b \left \{ a'' +{2ab''\over b} + 
{3a'b'\over b} + 2(D-2){1\over r}(a' +{ ab'
\over b}) + (D-2)(D-3){1\over r^2}(a-1) \right \}^2. 
\label{rsquare}
\ee
$D= 4$ is special, since the action is scale invariant and allows 
the following solution
\be
a(r) = 1 + {2M \over r} + {q^2 \over r^2 } + \Lambda r^2, \hskip 1 cm 
b(r) = c, 
\ee
with the interesting constraint
\be
q^2 \Lambda = 0. 
\ee 
Therefore, as in $C^2$ theory, there is a nice complementarity here: 
Either a charge or a 
cosmological term can be present, but not both.

For generic $D$, the equations derived from (\ref{rsquare}) 
allow only the RN  solution  
\be
a(r) = 1 + {2M \over r} + {q^2 \over r^{2D-6}},
\ee
since scale invariance is lost in $D \ne 4$.

Let us go back to $D=4$ and study another 'special' quadratic model in 
$D=4$. 
\be
I=  \int d^4x \sqrt{-g} {\tilde {R}}_{\mu \nu}{\tilde {R}}^{\mu \nu} 
\label{traceless}.
\ee 
where ${\tilde{R}}^{\mu \nu} = R_{\mu \nu} - {1\over 4}g_{\mu \nu}R $. 
What makes this model unique among 
the quadratic ones is that it is the only one whose solutions
have vanishing energy \cite{tekin}. 
[ From the Gauss-Bonnet identity, this model 
(\ref{traceless}) is equivalent to  
$\int d^4x \sqrt{-g} R_{\mu \nu \alpha \beta}   
R^{\mu \nu \alpha \beta}$ theory. Beyond $D=4$, the `traceless' theory 
$\int d^4x \sqrt{-g}\left \{ R_{\mu \nu} -{1\over D}
g_{\mu \nu} R \right \}^2 $ keeps its `zero-energy' property, but 
$\int d^4x \sqrt{-g} R_{\mu \nu \alpha \beta}^2$ does not.]     
The reduced action, not necessarily in the simplest form, is
\be
I = \int_0^\infty dr &\{& { 4 b \over r^2} ( a - 1 )^2 + 
{r^2 \over b}( 2ab'' + 3a'b')^2 +  4r^2aa''b'' \\ \nonumber
&& + 6r^2a'b'a'' + r^2ba''^2 + 8{a^2 b'^2\over b} + 8aa'b' + 4ba'^2 \}. 
\label{traceless2}
\ee 
After varying with respect to $(a, b)$, and setting $b = c$, one obtains 
two nonlinear equations satisfied by $a$, one of which is easily integrable 
with 4 integration constants. But the other equation requires 2 of those 
constants to be zero and one is left with {\it{only}} SdS.
While the fact that SdS is {\it{a}} solution of 
the model (\ref{traceless}) is also easy to see without 
using the Weyl trick, the latter allowed us 
to show that among the solutions of the 
type $-g_{00} = g^{rr}$, SdS is the {\it{only}} one!  
One could of course remove this restriction and try to solve the 
two highly non-linear equations in $(a,b)$. We do not know any exact 
solutions of this kind~\footnote{ Approximate solutions 
 were found (not via Weyl trick) for 
$R +  \alpha R_{\mu \nu \alpha \beta}^2$ in \cite{callan} } 
but the reduced action (\ref{traceless2}) is a good starting 
point in looking for such geometries.

\section{Axial Symmetry}

As the history of solutions in GR shows, things rapidly 
become more complicated with decreasing Killing directions :
In $D=4$, the evolution from Schwarzschild to Kerr  took some 45 years. 
For now, we just consider the simpler $D=3$ case, to  
obtain  its (well-known)  Kerr \cite{jackiw} or AdS  
Kerr \cite{btz} solutions. 
Here we take the time-dependent axially symmetric metric to be 
\be
ds^2 = - a(r,t)dt^2 + b(r,t)dr^2 + 2k(r,t)\,dt\,d\phi + h(r,t)d\phi^2. 
\ee
Birkhoff's theorem again easily emerges in this 
``spin 1'' world -there are no explicit time 
derivatives in the actions, even though the Ricci tensor and scalar have 
many time dependent terms. As a result, the action, with a cosmological 
constant, reduces to 
\be
I =  \int_{0}^{\infty} dr \left\{ { a'h' + k'^2 \over \sqrt{ abh + b k^2}} + 
\Lambda \sqrt{ abh + b k^2} \right\} 
\ee
up to boundary terms
\be
\Delta I = \int dr B' \hskip 1 cm  
B = - {(fh +k^2 )'\over \sqrt{ abh + b k^2} }.
\ee
We again fix the Schwarzschild gauge by setting  $h(r) = r^2$.
Then, varying the three remaining functions yields the solution
\be
k = -J, \hskip 1 cm a(r) = -c -\Lambda r^2 \hskip 1 cm b^{-1}(r) = 
- c -\Lambda r^2 + {J^2 \over 4 r^2}.  
\ee
The $\Lambda=0$ limit 
is the extension ( in slightly different coordinates ) of the original 
massless spinning metric of 
\cite{jackiw}, to include mass;   
if we keep $\Lambda$, we get the BTZ metric \cite{btz}.

\section{Summary}
We have attempted to highlight the Weyl method, 
within its legitimate scope, as a useful tool both 
for probing general properties such as Birkhoff's theorem 
and obtaining explicit solutions 
in the highly symmetric cases where they are 
plausibly available in general gravity models. Our explicit 
examples were primarily spherically, but also axially ,symmetric 
worlds ; the latter were also easily shown to exhibit time-independence.
Other, as yet unexplored, possible applications include 
brane-metrics, and non-abelian gauge theories.

We thank R. S. Palais for patient  discussions of his work and N. 
Wyllard for algebraic assistance. This research was supported by 
NSF Grant 99-73935



\begin{thebibliography}{99}

\bibitem{weyl}
H.~Weyl, ``Space-time-matter'', New York: Dover 1951.

\bibitem{palais}
R.~S.~Palais, 
``The Principle of Symmetric Criticality,'' 
Comm. \ Math. \ Physics {\bf{69}}, 19  (1979). 

\bibitem{torre}
M.~E.~Fels and C.~G.~Torre,
``The principle of symmetric criticality in general relativity,''
Class.\ Quant.\ Grav.\  {\bf {19}}, 641 (2002).
[arXiv:gr-qc/0108033].

\bibitem{jackiw}
S.~Deser, R.~Jackiw and G.~'t Hooft,
``Three-Dimensional Einstein Gravity: Dynamics Of Flat Space,''
Annals Phys.\  {\bf 152}, 220 (1984).


\bibitem{btz}
M.~Banados, C.~Teitelboim and J.~Zanelli,
``The Black Hole In Three-Dimensional Space-Time,''
Phys.\ Rev.\ Lett.\  {\bf 69}, 1849 (1992)
[arXiv:hep-th/9204099].



\bibitem{boulware}
D.~G.~Boulware and S.~Deser,
``String Generated Gravity Models,''
Phys.\ Rev.\ Lett.\  {\bf 55}, 2656 (1985).


\bibitem{riegert}
R.~J.~Riegert,
``Birkhoff's Theorem in Conformal Gravity'',
Phys.\ Rev.\ Lett. {\bf 53}, 315-318 (1984).


\bibitem{mannheim}
P.~D.~Mannheim and D.~Kazanas,
``Exact Vacuum Solution To Conformal
Weyl Gravity And Galactic Rotation Curves,''
Astrophys.\ J.\  {\bf 342}, 635 (1989).


\bibitem{tekin}
S.~Deser and B.~Tekin,
``Gravitational energy in quadratic curvature gravities,''
Phys.\ Rev.\ Lett.\  {\bf 89}, 101101 (2002)
[arXiv:hep-th/0205318]; ``Energy in generic higher curvature 
gravity theories,''
Phys.\ Rev.\ D {\bf 67}, 084009 (2003)
[arXiv:hep-th/0212292].


\bibitem{callan}
C.~G.~Callan, R.~C.~Myers and M.~J.~Perry,
``Black Holes In String Theory,''
Nucl.\ Phys.\ B {\bf 311}, 673 (1989).



\end{thebibliography}
\end{document}